\documentclass[12pt, journal,draftclsnofoot,a4paper,oneside,onecolumn]{IEEEtran}
\usepackage{amsfonts}
\usepackage{mathrsfs}

\usepackage{graphicx}

\ifCLASSINFOpdf
\else
\fi
\begin{document}

\title{On the Minimum Differential Feedback for Time-Correlated MIMO Rayleigh Block-Fading Channels}

\author{\IEEEauthorblockN{Leiming Zhang, Lingyang Song, Meng Ma, and Bingli Jiao}
\IEEEauthorblockA{\\School of Electrical Engineering and Computer
Science\\ Peking University, Beijing, China $100871$\\
Email: \{leiming.zhang, lingyang.song, mam, jiaobl\}@pku.edu.cn.\\
}}



%


\maketitle

\begin{abstract}
In this paper, we consider a general multiple input multiple output~(MIMO) system with channel state information (CSI) feedback over time-correlated Rayleigh block-fading channels. Specifically, we first derive the closed-form expression of the minimum differential feedback rate to achieve the maximum erdodic capacity in the presence of channel estimation errors and quantization distortion at the receiver. With the feedback-channel transmission rate constraint, in the periodic feedback system, we further investigate the relationship of the ergodic capacity and the differential feedback interval, and we find by theoretical analysis that there exists an optimal differential feedback interval to maximize ergodic capacity. Finally, analytical results are verified through simulations in a practical periodic differential feedback system using Lloyd's quantization algorithm.
\end{abstract}


%
\IEEEpeerreviewmaketitle

\section{Introduction}
Channel state information (CSI) feedback from the receiver to the transmitter has been intensively studied with great interest due to its potential benefits to the multiple input multiple output (MIMO) system. CSI can be utilized by a variety of channel adaptive techniques (e.g., water-filling, beamforming, precoding, etc.) at the transmitter to enhance the spectral efficiency as well as the robustness of the system, especially, in the frequency division duplexing (FDD) mode. As the transmission rate of the feedback channel is normally very limited, the infinite feedback of CSI is hard to realize in practice. Therefore, it is important to investigate how to decrease the amount of feedback signalling overhead to meet the uplink feedback channel requirements. As a result, CSI feedback reduction has attracted lots of attention in recent years~\cite{Ref:David-ovwerview}, \cite {Ref:David-value of feedback}.

Specifically, when the wireless channel experiences time-correlated fading \cite{Ref:Jack's model}, typically represented by a Markov random process \cite{Ref:K.Huang-Markov models}--\cite{Ref:K.E.Baddour-AR model}, the amount of CSI feedback can be largely reduced. In \cite{Ref:T.Eriksson-summary of compression}, a number of feedback reduction schemes were summarized, considering the lossy compression scheme exploiting the properties of fading process as the best choice. In \cite{B.Mondal-switched codebook} and \cite{Ref:T.Kim-differential
rotation feedback}, schemes using switched codebook and rotation codebook with differential feedback  were proposed, respectively. In \cite{Ref:K.Huang-Markov compression}, it modeled the time-correlated fading channel as a finite-state Markov chain to reduce the feedback rate by ignoring some states occurred with small probabilities. In \cite{Ref:C.Simon-feedback compression} and \cite{K.Kim-utilizing emporal correlation}, a predictive vector quantization scheme was proposed, provided that the previous quantization CSI is known. In \cite{Ref:C.Simon-Variable}, variable-length code was applied for feedback rate reduction. Despite of so much research on practical feedback reduction schemes, the lower bound of the feedback compression as well as the required minimum differential feedback rate to guarantee the accuracy of CSI has not yet been well studied for time-correlated MIMO Rayleigh block-fading channels.

In \cite{Ref: D.Zhang-capacity and feedback rate}--\cite{Ref:A.D.Dabbagh-capacity loss}, the relationship between the capacity gain and the limited feedback of CSI was studied. Lower and upper bounds of ergodic capacity gain using CSI feedback in comparison with open-loop systems were reported in \cite{Ref:D.Zhang-bounds of feedback rate} and \cite{Ref:W.Li-comments on}. However, the time correlation was not taken into account in these work. In \cite{Ref:W.H.Li-adaptive mimo transmission over time vary channels}, a periodic feedback scheme was studied with time correlation, but the feedback only occurs in the first block of the transmission period. Unlike previous work, in this paper, we investigate the relationship between the ergodic capacity and the differential feedback interval with feedback channel capacity constraint in every fading block. 

In  this paper, we consider a general MIMO system with periodic differential CSI feedback over time-correlated Rayleigh block-fading channels, and address the problem of the ergodic capacity under the impact of the feedback interval. The main contribution can be briefly summarized as follows:
\begin{enumerate}
  \item We derive the minimum differential feedback rate for time-correlated MIMO Rayleigh block-fading channels by taking into account of both the channel estimation errors and channel quantization distortion.
  \item We investigate the relationship between the ergodic capacity and the differential feedback interval with feedback channel rate constraint in a periodic
feedback system. Furthermore, we prove that there exists an optimal feedback interval to achieve the maximum ergodic capacity.
  \item We design a practical differential feedback scheme using Lloyd's quantization algorithm to verify the theoretical results.
\end{enumerate}

The rest of the paper is organized as follows. In Section~II, we describe the system model. In Section~III, the minimum differential feedback rate is derived, and the relationship between the ergodic capacity and differential feedback interval is studied. In Section~IV, we provide the simulation
results. In Section~V, we draw the main conclusions. The derivations
are given in the appendices.

\emph{\textbf{{Notation:}}} Bold uppercase (lowercase) letters denote matrices (vectors), ${(\cdot)^ +}$ denotes Hermitian
transpose, $\log_2(\cdot)$ denotes the base two logarithm,
$\det(\cdot)$ denotes determinant operator, and $\mathbb{E}[\cdot]$
stands for the expectation over random variables.


\section{System Model}
\begin{figure}[]
\centering
\includegraphics[width=6in]{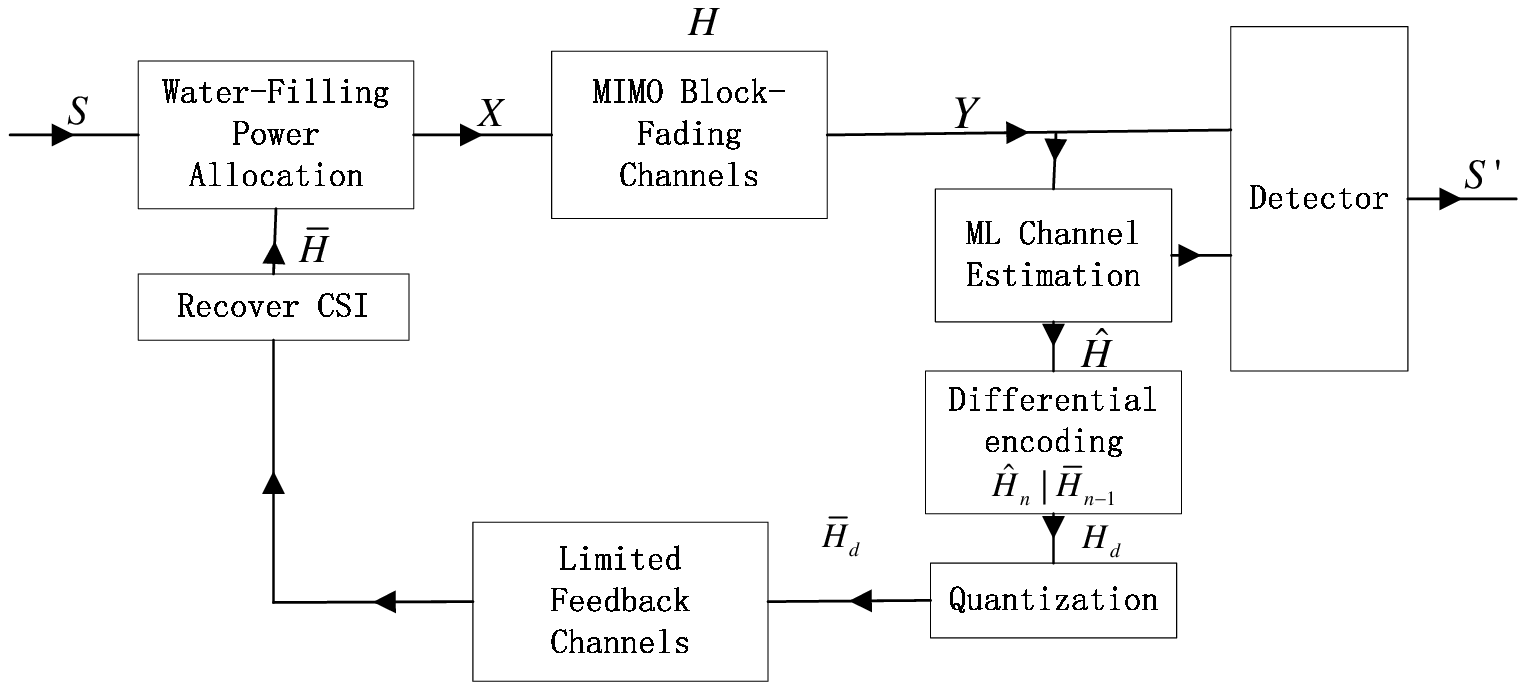}
\caption{System Model.} \label{fig:1}
\end{figure}

The system model is illustrated in Fig.~\ref{fig:1}, where the downlink channel is modeled as a time-correlated MIMO Rayleigh block-fading channels, and the uplink channel is modeled as a limited and lossless feedback channel with a feedback capacity
constraint per fading block.
In this paper, we consider the differential feedback, i.e., the receiver just feeds back the differential CSI to the transmitter given the previous channel quantization matrix, where the channel estimation errors and channel quantization distortion are also considered.

\subsection{Time-Correlated MIMO Rayleigh Block-Fading Channel Model}

We consider MIMO Rayleigh block fading channels, where the channel fading matrix remains constant within a fading block and varies from one to another. There are
$N_t$ transmitter antennas and $N_r$ receiver antennas. The received signals can be expressed in a vector form
\begin{equation}\label{eq:1}
\mathbf{y} = \mathbf{Hx} + {\mathbf{n}_0},
\end{equation}
where ${\bf{y}} = {[{y_1},{y_2},{y_3}, \ldots ,{y_{{N_r}}}]^T}$ denotes a ${N_r} \times 1$ received signal vector, $\mathbf{H}$ is a ${N_r} \times {N_t}$ channel fading matrix with independent entries obeying complex Gaussian distribution $\mathcal{C}\mathcal {N}(0,\sigma _h^2)$, ${\bf{x}} = {[{x_1},{x_2},{x_3},
\ldots ,{x_{{N_t}}}]^T}$ represents a ${N_t} \times 1$ transmitted
signal vector, and $\mathbf{n}_0$ is a ${N_r} \times 1$ noise vector
whose entries are independent and identically distributed (i.i.d)
complex Gaussian variables $\mathcal {C}\mathcal {N}(0,\sigma
_0^2)$.

The time-correlated channel can be represented by a first-order Autoregressive model~(AR1)~\cite{Ref:K.E.Baddour-AR model}, and the channel fading matrix can be
written as
\begin{equation}\label{eq:2}
{\mathbf{H}_n} = \alpha {\mathbf{H}_{n - 1}} + \sqrt {1 - {\alpha
^2}} {\mathbf{W}_n},
\end{equation}
where ${\mathbf{H}_{n - 1}}$ denotes ${(n - 1)_{th}}$ channel fading
matrix, $\mathbf{W}_n$ is a noise matrix, which is independent of
${\mathbf{H}_{n - 1}}$, and the entries are i.i.d. complex Gaussian
variables $\mathcal {C}\mathcal {N}(0,\sigma _h^2)$. The parameter
$\alpha$ is the time autocorrelation coefficient, which is given by
the zero-order Bessel function of first kind $\alpha  = {J_0}(2\pi
{f_d}\tau )$ \cite{Ref:K.E.Baddour-AR model}, where $f_d$ denotes
the maximum Doppler frequency in Hertz, and $\tau$ denotes the time
interval. In the block-fading system, the time interval can be
calculated as $T = \tau /{t_{block}}$, where $t_{block}$ is the
duration of every block.

The CSI can be estimated by the receiver using orthogonal pilots. Without loss of generality, in this paper, maximum likelihood
(ML) criterion is employed for channel estimation, and the estimated channel matrix can be expressed in an equivalent form as
\begin{equation}\label{eq:3}
\hat \mathbf{H} = \mathbf{H} + \mathbf{H}_e,
\end{equation}
where $\hat \mathbf{H}$ denotes the channel estimation matrix, whose entries are i.i.d. complex Gaussian variables
$\mathcal {C}\mathcal {N}(0,\sigma _{\hat h}^2)$, $\mathbf{H}$ is the actual channel fading matrix, and $\bf{H}_e$ denotes
the channel estimation error matrix, which is independent of
$\mathbf{H}$, with entries of independent complex Gaussian
distributed with $\mathcal {C}\mathcal {N}(0,\sigma _{\hat h}^2 -
\sigma _h^2)$~\cite{Ref:D.Sam-Pilot assisted
estimation}.

As $\mathbf{H}_e$ is independent of $\mathbf{H}$ in (\ref {eq:3}),
we obtain
\begin{equation}\label{eq:estimation-2}
{\mathbf{H}} = \frac{{\sigma _h^2}}{{\sigma _{\hat
h}^2}}\hat{\mathbf{H}} + \bf{\Psi} ,
\end{equation}
where $\sigma _h^2$ and $\sigma _{\hat h}^2$ denote the variances of
$\mathbf{H}$ and $\hat \mathbf{H}$, respectively, and $\bf{\Psi}$ is
independent of $\hat \mathbf{H}$ with entries satisfying $\mathcal
{C}\mathcal {N}\left( {0,\frac{{\sigma _h^2 \cdot (\sigma _{\hat
h}^2 - \sigma _h^2)}}{{\sigma _{\hat h}^2}}} \right)$. The detailed
derivation of (\ref{eq:estimation-2}) is given in Appendix~A.

\subsection{CSI Feedback Model}

We consider a limited and lossless feedback channel. Through CSI
quantization, the feedback channel output $\bar \mathbf{H}$ can be
modeled as \cite{Ref:D.Zhang-bounds of feedback rate}
\begin{equation}\label{eq:CSI quantization}
\hat \mathbf{H} = \bar \mathbf{H} + \mathbf{E},
\end{equation}
where $\mathbf{E}$ denotes the independent additive quantization error matrix with entries satisfying $\mathcal {C}\mathcal
{N}\left( {0,\frac{D}{{{N_r}{N_t}}}} \right)$, where $D$ represents
the channel quantization distortion constraint.

In this paper, we consider the differential feedback, where only the differential CSI will be fed back to the transmitter, assuming that the previous channel quantization matrix ${\bar \mathbf{H}_{n - 1}}$ is known both at receiver and transmitter. The differential CSI can be
written as
\begin{equation}\label{eq:6}
{\mathbf{H}_d} = Diff\left( {\hat \mathbf{H}_n, {{\bar
\mathbf{H}}_{n - 1}}} \right),
\end{equation}
where ${\mathbf{H}_d}$ represents the differential CSI between $\hat
\mathbf{H}_n$ and ${\bar \mathbf{H}_{n - 1}}$, and $Diff( \cdot )$
denotes the differential function. 

Furthermore, we assume that the
CSI feedback channel has a capacity constraint ${C_{fb}}$ per fading
block. When the CSI is quantized to $R$ bits and the feedback
interval is $T$ blocks, the average feedback rate satisfies the
inequality $R/T \le {C_{fb}}$. Therefore, the feedback interval can
be calculated by
\begin{equation}\label{eq:feedback interval}
T = \left\lceil {\frac{R}{{{C_{fb}}}}} \right\rceil,
\end{equation}
where $\left\lceil x \right\rceil$ denotes the smallest integer
larger than $x$.


\subsection{Ergodic Capacity of Pilot-assisted MIMO Systems}
In this paper, we use the water-filling precoder to obtain the
capacity gain. The channel quantization matrix can be
decomposed at the transmitter to perform water-filling ??????water-filling reference??????
\begin{equation}\label{eq:8}
\bar \mathbf{H} = \mathbf{U}\bf{\Sigma} {\mathbf{V}^ + },
\end{equation}
where $\mathbf{U}$ and $\mathbf{V}$ are unitary matrixes, and
$\bf{\Sigma}$ is a non-negative and diagonal matrix composed of
eigenvalues.

For the pilot-assisted MIMO system with ML channel estimation, the
closed-loop ergodic capacity with water-filling can be obtained with
the help of \cite{Ref:W.H.Li-adaptive mimo transmission over time
vary channels,Ref:D.Sam-Pilot assisted estimation}
\begin{equation}\label{eq:9}
{C_{erg}} = \mathbb{E}_{\hat \mathbf{H},\bar \mathbf{H}}\left[
{\frac{{L - {N_t}}}{L}{{\log }_2}\det \left( {{\mathbf{I}_{{N_r}}} +
\mathbf{J}\cdot{\mathbf{J}^ + }\left( {{\mathbf{F}^{ - 1}}} \right)}
\right)} \right],
\end{equation}
where $\mathbf{J} = \hat \mathbf{H}\mathbf{V}\mathbf{Z}$,
${\mathbf{J}_e} = {\mathbf{H}_e}\mathbf{V}\mathbf{Z}$, $\mathbf{F} = \frac{1}{A^2}{\mathbf{I}_{{N_r}}} +
{E_{{\mathbf{J}_e}}}\left[ {{\mathbf{J}_e}\mathbf{J}_e^ +
|\mathbf{J}} \right]$, $L$ denotes the number of transmitted
symbols, $A$ represents the amplitude of signal symbol, and
$\mathbf{Z}$ stands for a diagonal matrix determined by the
water-filling algorithm, which is given by ??????reference??????
\begin{equation}\label{eq:waterfilling}
\left\{ \begin{array}{l}
 z_{_i}^2 = \left\{ \begin{array}{l}
 \mu  - {\left( {\gamma _{i,i}^2{A^2}} \right)^{ - 1}},\;\gamma _{i,i}^2{A^2} \ge {\mu ^{ - 1}} \\
 0,\quad \quad \quad \quad \quad \;otherwise \\
 \end{array} \right. \\
 \sum\limits_{i = 1}^{{N_t}} {z_{_i}^2{A^2} = {N_t}{A^2},}  \\
 \end{array} \right.
\end{equation}
where ${\gamma _{i,i}}$ are entries of $\mathbf{\Sigma}$, and $\mu$
is a cut-off value chosen to meet the power constraint.

It can be observed from (\ref{eq:9}) that the closed-loop ergodic capacity is determined by $\bar \mathbf{H}$ and $\hat \mathbf{H}$, and the loss of the capacity is mainly caused by the distortion. Hence, the ergodic capacity is a negative-correlated function in association with the distortion of CSI feedback~\cite{Ref: D.Zhang-capacity and feedback rate,Ref:A.D.Dabbagh-capacity loss}.


\section{Minimum Differential Feedback Rate}
In this section, we derive the minimum differential feedback rate of
the time-correlated MIMO Rayleigh block-fading channels to guarantee
the accuracy of the CSI. The minimum differential feedback rate is
determined by the rate distortion theory of continuous-amplitude
sources ??????reference??????. When the ${(n - 1)_{th}}$ channel quantization matrix
${\bar \mathbf{H}_{n - 1}}$ is known at both receiver and
transmitter, the minimum differential feedback rate can be written
as
\begin{equation}\label{eq:10}
R = \inf \left\{ {I\left( {{{\hat \mathbf{H}}_n};{{\bar
\mathbf{H}}_n}|{{\bar \mathbf{H}}_{n - 1}}} \right):E\left[ {d\left(
{{{\hat \mathbf{H}}_n};{{\bar \mathbf{H}}_n}} \right)} \right] \le
D} \right\},
\end{equation}
where $\inf \left\{  \cdot  \right\}$ denotes infimum function
\cite{Ref:R.J-rate-distotion}, ${I\left(
{{{\hat{\mathbf{H}}}_n};{{\overline {\mathbf{H}} }_n}|{{\overline
{\mathbf{H}} }_{n - 1}}} \right)}$ denotes the mutual information
between $\hat \mathbf{H}_n$ and $\bar \mathbf{H}_n$ given $\bar
\mathbf{H}_{n-1}$, and $d\left( {{{\hat \mathbf{H}}_n},{{\bar
\mathbf{H}}_n}} \right) = {\left\| {{{\hat \mathbf{H}}_n} - {{\bar
\mathbf{H}}_n}} \right\|^2}$ is the channel quantization distortion,
which is the measurement of the quality of feedback information.

Since the entries of $\mathbf{H}$, $\hat \mathbf{H}$ and $\bar
\mathbf{H}$ are i.i.d. complex Gaussian variables, the minimum
differential feedback rate can be written as
\begin{equation}\label{eq:mutual information}
R= \inf \left\{ {{N_r}{N_t} \cdot I\left( {{{\hat h}_n};{{\bar
h}_n}|{{\bar h}_{n - 1}}} \right):E\left[ {d\left( {{{\hat
h}_n};{{\bar h}_n}} \right)} \right] \le d} \right\},
\end{equation}
where $d = \frac{D}{{{N_r}{N_t}}}$ denotes the one-dimensional
average channel quantization distortion, and ${\hat h_n}$, ${\bar h_n}$,
and ${\bar h_{n - 1}}$ denote the entries of ${\hat \mathbf{H}_n}$,
${\bar \mathbf{H}_n}$, and ${\bar \mathbf{H}_{n - 1}}$,
respectively.

\newtheorem{Lemma}{Lemma}
\begin{Lemma}\label {lemma:1}
Given the one-dimensional channel quantization distortion constraint
$d$, and the ${\left( {n - 1} \right)_{th}}$ channel quantization
element ${\bar h_{n - 1}}$, the mutual information $I\left( {{{\hat
h}_n};{{\bar h}_n}|{{\bar h}_{n - 1}}} \right)$ can be calculated as
\begin{equation}\label{eq:lemma1}
I \ge \log \left[ {{\alpha ^2}{{\left( {\frac{{\sigma _h^2}}{{\sigma
_{\hat h}^2}}} \right)}^2} + \frac{{\left( {1 - {\alpha ^2}}
\right)}}{d}\sigma _h^2 + \frac{{\left( {\sigma _{\hat h}^2 - \sigma
_h^2} \right)}}{d}} \right. \left. {\left( {1 + {\alpha
^2}\frac{{\sigma _h^2}}{{\sigma _{\hat h}^2}}} \right)} \right],
\end{equation}
where $\sigma _h^2$ and $\sigma _{\hat h}^2$ denote the variances of
$h$ and $\hat h$ respectively, and $\alpha$ is the time
autocorrelation coefficient.
\end{Lemma}

The proof of \emph{Lemma 1} can be found in Appendix B. As
${\hat h_n}$, ${\bar h_n}$ and ${\bar h_{n - 1}}$ are complex
Gaussian variables, the minimum value of the mutual information is
indeed achievable \cite{Ref:R.J-rate-distotion}.

Combining (\ref {eq:mutual information}) and (\ref {eq:lemma1}), the
minimum differential feedback rate of the time-correlated MIMO
block-fading channels can be calculated as

\begin{equation}\label{eq:minimum rate}
R= {N_r}{N_t}\cdot\max \left\{ {\log \left[ {{\alpha ^2}{{\left(
{\frac{{\sigma _h^2}}{{\sigma _{\hat h}^2}}} \right)}^2} +
\frac{{\left( {1 - {\alpha ^2}} \right)}}{d}\sigma _h^2} \right.}
\right.\left.{\left. { + \frac{{\left( {\sigma _{\hat h}^2 - \sigma
_h^2} \right)}}{d}\left( {1 + {\alpha ^2}\frac{{\sigma
_h^2}}{{\sigma _{\hat h}^2}}} \right)} \right],0} \right\}.
\end{equation}
From (\ref{eq:minimum rate}), we can see that the minimum
differential feedback rate is determined by the distortion of the
quantization, time correlation coefficient, and the estimation
variance. Note that the minimum differential feedback rate in
(\ref{eq:minimum rate}) is the lower bound of feedback compression
with time correlation in the block-fading MIMO channels. Given the
accuracy of feedback CSI (i.e. the distortion $d$), the minimum
feedback rate can be easily obtained in (\ref{eq:minimum rate}).

Furthermore, as the ergodic capacity increases with the distortion
decreasing, we investigate the feedback design scheme for minimizing
the distortion of the feedback CSI in order to maximize the ergodic
capacity in the following.

From (\ref{eq:minimum rate}), if $R \ge 0$, $d$ can be calculated as
\begin{equation}\label{eq:14}
d = \left( {\sigma _{\hat h}^2 - {{\left( {\frac{{\sigma
_h^2}}{{\sigma _{\hat h}^2}}} \right)}^2}\sigma _{\hat
h}^2\cdot{\alpha ^2}} \right)/\left( {{2^{\frac{R}{{{N_r}{N_t}}}}} -
{\alpha ^2}{{\left( {\frac{{\sigma _h^2}}{{\sigma _{\hat h}^2}}}
\right)}^2}} \right).
\end{equation}
In a practical communication system, the feedback channel is causal,
which implies that ${\bar \mathbf{H}_n}$ can be only used in the
next feedback period ${{{\hat{\mathbf{H}}}_{n + 1}}}$. With the
causal feedback constraint, we consider the impact of the feedback
delay on the distortion. Combining (\ref {eq:14}) and (\ref
{eq:30}), the distortion can be written as
\begin{equation}\label{eq:15}
d = {\alpha ^2}{\left( {\frac{{\sigma _h^2}}{{\sigma _{\hat h}^2}}}
\right)^2}\frac{{\sigma _{\hat h}^2 - {{\left( {\frac{{\sigma
_h^2}}{{\sigma _{\hat h}^2}}} \right)}^2}\sigma _{\hat h}^2 \cdot
{\alpha ^2}}}{{{2^{\frac{R}{{{N_r}{N_t}}}}} - {\alpha ^2}{{\left(
{\frac{{\sigma _h^2}}{{\sigma _{\hat h}^2}}} \right)}^2}}} + {\alpha
^2}\frac{{\sigma _h^2\left( {\sigma _{\hat h}^2 - \sigma _h^2}
\right)}}{{\sigma _{\hat h}^2}}+ \left( {1 - {\alpha ^2}}
\right)\sigma _h^2 + \left( {\sigma _{\hat h}^2 - \sigma _h^2}
\right).
\end{equation}


Given $\sigma _h^2$ and $\sigma _{\hat h}^2$, we can see that $d$ is
a function of $R$ and $\alpha$ in (\ref {eq:15}). In a periodic
feedback system with limited feedback, indicated by (\ref{eq:2}) and
(\ref{eq:feedback interval}), both $\alpha$ and $R$ are related to
$T$. Therefore, after some manipulations, the distortion $d$ can be
expressed as a function of $T$,
\begin{equation}\label{eq:D-T}
d\left( T \right) = \frac{{\sigma _h^4}}{{\sigma _{\hat
h}^2}}\cdot\left( {\frac{{\left( {1 -
{2^{\frac{{{C_{fb}}T}}{{{N_r}{N_t}}}}}} \right)\alpha {{\left( T
\right)}^2}}}{{{2^{\frac{{{C_{fb}}T}}{{{N_r}{N_t}}}}} - {{\left(
{\frac{{\sigma _h^2}}{{\sigma _{\hat h}^2}}} \right)}^2}\alpha
{{\left( T \right)}^2}}}} \right) + \sigma _{\hat h}^2.
\end{equation}
 From (17), we have
\begin{equation}\label{eq:zero}
T \to 0\qquad  \Rightarrow \quad
\quad{2^{\frac{{{C_{fb}}T}}{{{N_r}{N_t}}}}} \to 1\qquad \Rightarrow
\quad d \to \sigma _{\hat h}^2.
\end{equation}
Similarly, when $T$ is large enough, the time correlation
$\alpha\left( T \right)$ trends to $0$. Therefore, we have
\begin{equation}\label{eq:infty}
T \to \infty \qquad  \Rightarrow \quad \alpha (T) \to 0\qquad
\Rightarrow \quad d \to \sigma _{\hat h}^2.
\end{equation}
There are some interesting observations from (18) and (19): When $T$ trends to
zero, the channel state remains static, such that it is not necessary to send any feedback bits. Therefore, the quantization channel at the transmitter is independent of the estimation channels at receiver. On the other hand, if $T$ is large enough, the time
correlation decreases to zero, which implies that the feedback quantization channel is completely outdated and it is also independent
of the estimation channel. Therefore, the distortion in both (18)
and (19) are $\sigma _{\hat h}^2$.

\begin{figure}[]
\centering
\includegraphics[width=6in]{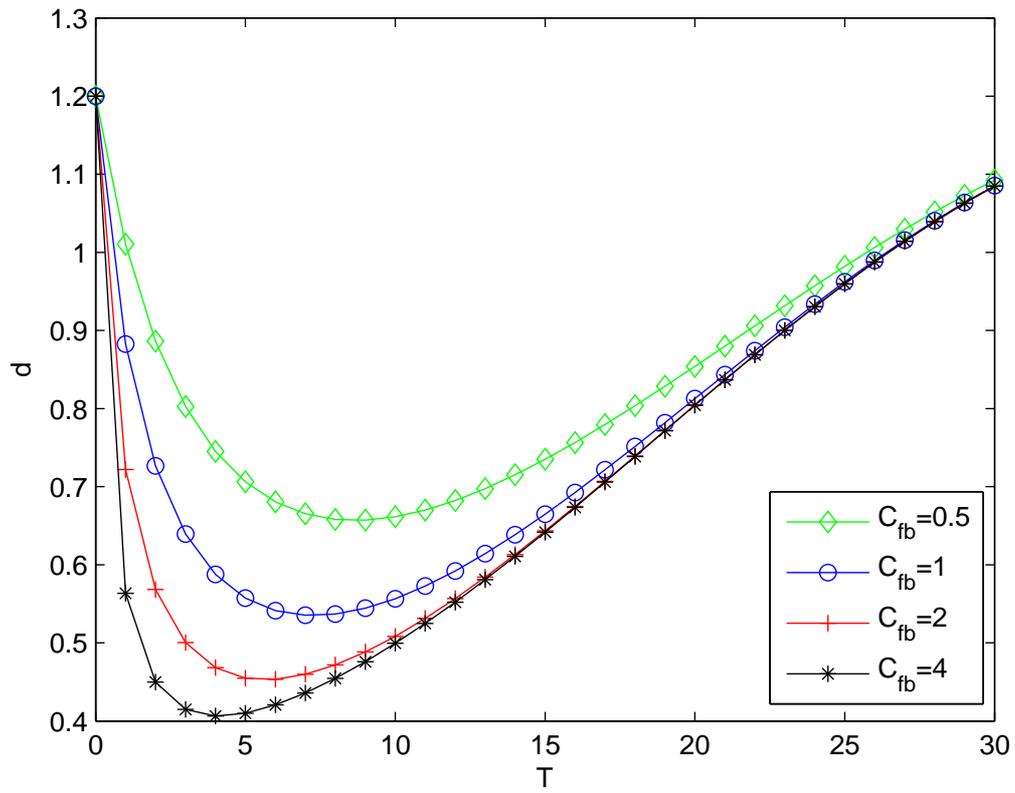}
\caption{The relationship between the distortion of channel state information feedback and the feedback interval for $N_r=2$, $N_t=2$,
${\sigma _h^2}=1$ and ${\sigma _{\hat h}^2}=1.2$.} \label{fig:2}
\end{figure}

When $0 < T < \infty $, we have $0 < {\alpha}\left( T \right)^2 < 1$
and $1 < {2^{\frac{{{C_{fb}}T}}{{{N_r}{N_t}}}}}$. Hence, we can obtain that the first
term of (\ref{eq:D-T}) is negative, and
\begin{equation}\label{eq:inter}
0 < T < \infty  \qquad \Rightarrow \qquad d < \sigma _{\hat h}^2.
\end{equation}
Combining (\ref{eq:zero}), (\ref{eq:infty}) and (\ref{eq:inter}), we
can predict that there exists an optimal $T$ in the region
$(0,\infty )$ to minimize the distortion. We give the proof of the
existence of the optimal feedback interval $T$ in the Appendix C. To further verify the theoretical analysis, numerical results of the relationship between the
distortion and the feedback interval from (\ref{eq:D-T}) are given in Fig.~\ref{fig:2}. 




\section{Simulation Results and Discussion}

In this section, we first provide the simulation results for the
derived minimum differential feedback rate expression. Then, we
discuss the relations between the ergodic capacity andthe  feedback
interval in a periodic feedback system with feedback channel
transmission rate constraint. Finally, we verify our theoretical
results by a practical differential feedback system employing Lloyd's
quantization algorithm. All simulations are performed for a point-to-point MIMO system over time-correlated block fading channels. For simplicity and without loss of generality, we consider ${N_t} = 2$ antennas at transmitter, ${N_r} = 2$ antennas at receiver, and the channel variance is set as $\sigma _h^2 = 1$.

\subsection{Minimum Differential Feedback Rate}

\begin{figure}[]
\centering
\includegraphics[width=6in]{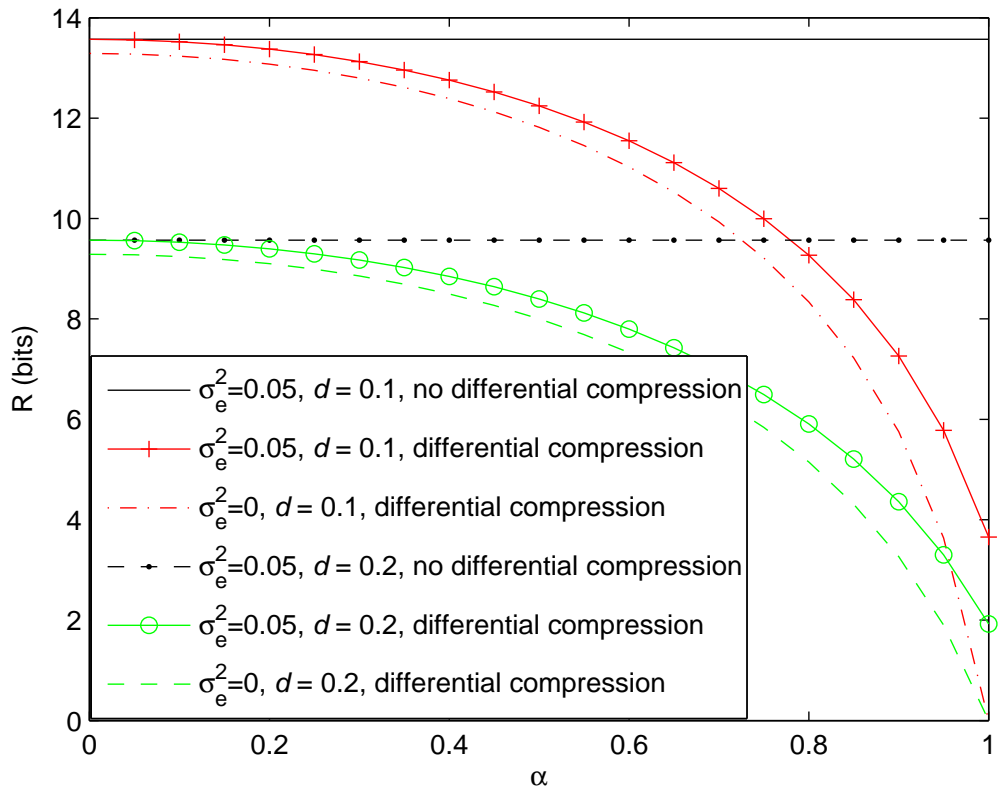}
\caption{The relationship between the minimum differential feedback
rate and time correlation for $N_r=2$, $N_t=2$, $\sigma _e^2  =
\left\{ {0, 0.05}\right\}$ and $d = \{ 0.1, 0.2\}$.} \label{fig:3}
\end{figure}

Fig.~\ref{fig:3} shows that the minimum differential feedback rate
versus the time correlation with the variance of channel estimation
error $\sigma _e^2 = \sigma _{\hat h}^2 - \sigma _h^2 = \left\{ {0,
0.05} \right\}$, and the accuracy of CSI is represented by the
distortion with $d = \{ 0.1, 0.2\}$. We also include the
non-differential compression results for comparison.

In Fig.~\ref{fig:3}, we can see that when time correlation increases,
it results in significant reduction of feedback rate by using
differential compression. In addition, the impact of estimation
error and quantization distortion is also illustrated in
Fig.~\ref{fig:3}. For lower quantization distortion, larger minimum
feedback rate is required. It can be also observed from
Fig.~\ref{fig:3} that with more estimation errors, the feedback rate has to be increased. 


\subsection{Ergodic Capacity and Feedback Interval}

In this subsection, we give the simulation results of the relationship
between the ergodic capacity and the feedback intervals. For simplicity, we assume that the block size is $L = 100$ with the duration of $1$ ms, and the power of pilot is $10\%$ of the
total transmit power, which is a reasonable value in practice
\cite{Ref:D.Sam-Pilot assisted estimation}. We select a relatively smaller
value of SNR, which is $0$ dB, and the Doppler frequency is $9.26$ Hz (Moving speed
is $5$ km/h, and the Carrier Frequency is $2$ GHz).

%
%
%
%

\begin{figure}[]
\centering
\includegraphics[width=6in]{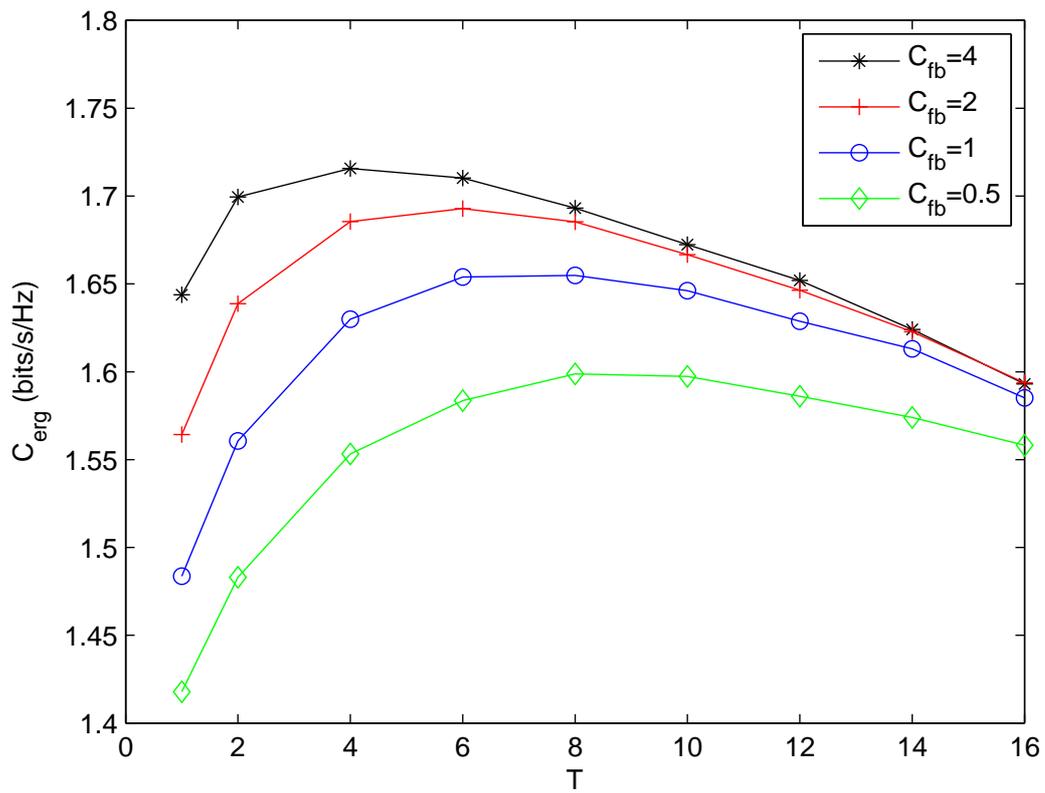}
\caption{The relationship between the ergodic capacity and feedback
interval for ${N_r} = 2,{N_t} = 2$, $SNR=0 dB$, $L=100$ and
${f_D}=9.26$ Hz.} \label{fig:4}
\end{figure}

In Fig.~\ref {fig:4}, we plot the relations between ergodic
capacity and the feedback interval with the feedback capacity constraint
$C_{fb}=\{ 0.5, 1, 2, 4\}$ for every block. It clearly shows that
the ergodic capacity is a monotonic convex function of the feedback
interval, and there exists an optimal feedback interval which maximizes the ergodic capacity. The results are reasonable, because
when $T$ increases from a small region, it begins to provide larger
feedback rate and thus improve the quality of feedback information,
while when $T$ goes toward a relatively larger region, the time
correlation gradually decreases and the feedback delay becomes
larger, causing the feedback information outdated and therefore
impair the performance.

Note that the relations between $C_{erg}$ and $T$ in Fig.~\ref
{fig:4} is consistent with the analysis in section III, and the
similar optimal values of $T$ can be also found in Fig.~\ref
{fig:2}. Additionally, from Fig.~\ref {fig:4}, we can see that as
${C_{fb}}$ increases, the ergodic capacity also enhances. However,
the absolute increment becomes smaller, which implies that it is
necessary to limit the feedback channel transmission rate since
little gain can be achieved when ${C_{fb}}$ becomes very large.


\subsection{Differential Feedback System with Lloyd's Quantization Algorithm}
In order to verify our theoretical results, we design a differential
feedback system using Lloyd's quantization algorithm
\cite{Ref:Lloyd-Lloyd Algerithm}. Firstly, differential codebooks
are generated by Lloyd's quantization algorithm and available at
both receiver and transmitter. When the ${\left( {n - 1}
\right)_{th}}$ channel quantization matrix ${\bar \mathbf{H}_{n -
1}}$ is known both at receiver and transmitter, the receiver only
feeds back the differential codeword to the transmitter.

The feedback steps are given as follows. Firstly, the receiver
calculates true quantization error ${\mathbf{H}_d} = {\hat
\mathbf{H}_n} - {\bar \mathbf{H}_{n - 1}}$. Secondly, this true
error is quantized as ${\mathbf{C}_d}$ in the differential codebooks
with the smallest Euclidean distance to the true error. Thirdly, the
corresponding codeword index is sent back to the transmitter.
Finally, the transmitter recovers the channel quantization matrix by
${\bar \mathbf{H}_n} = {\bar \mathbf{H}_{n - 1}} + {\mathbf{C}_d}$.

\begin{figure}[]
\centering
\includegraphics[width=6in]{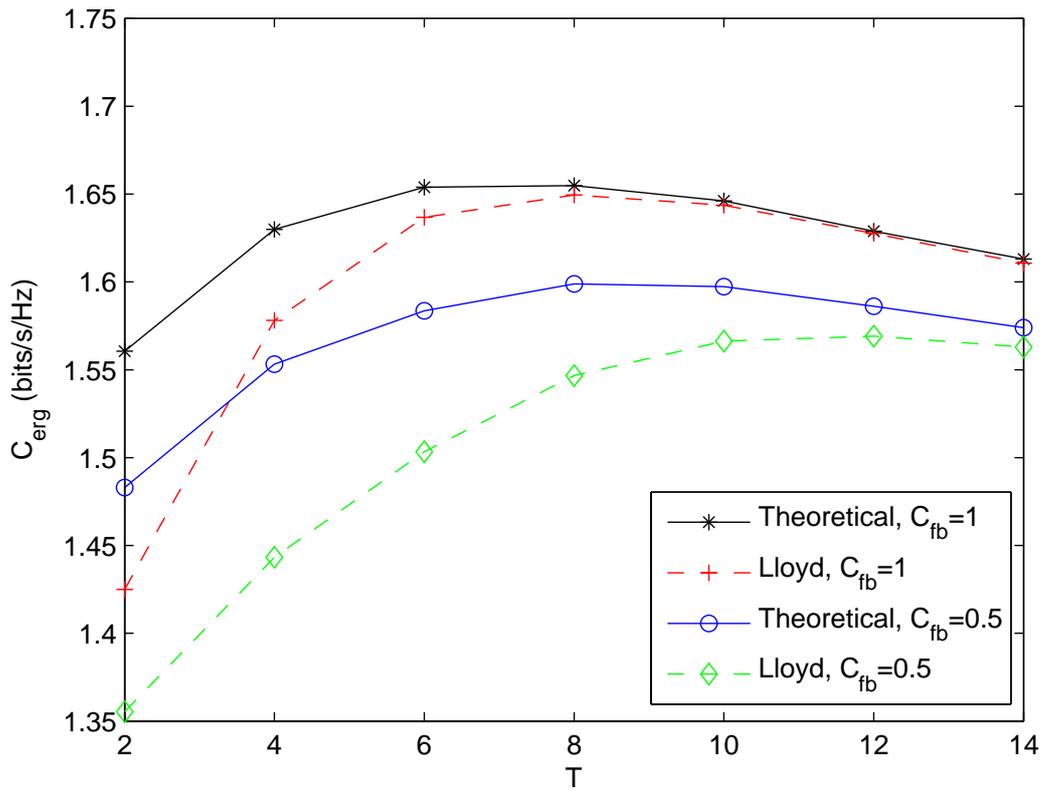}
\caption{The relationship between the ergodic capacity and feedback
interval with Lloyd algorithm for ${N_r} = 2,{N_t} = 2$, $SNR=0 dB$,
$L=100$ and ${f_D}=9.26$ Hz.} \label{fig:5}
\end{figure}

Fig.~\ref {fig:5} shows the ergodic capacity using the Lloyd's
quantization algorithm (dash curves) have the same trend with
theoretical ones (solid curves) and there exists an optimal feedback
interval. From Fig.~\ref {fig:5}, it shows that the ergodic capacity
of theoretical results is larger than the practical ones at small
feedback interval region, but they get converged as the feedback
interval increases. The reasons are given as follows. When the
feedback interval is in the small region, the feedback rate is not
sufficient both for theoretical and practical results, since the
codebooks generated with Lloyd's quantization algorithm have
stronger randomization. However, when the feedback rate is small,
with the increase of the feedback interval, the more feedback rate
can be obtained, reducing the randomization of Lloyd's quantization
algorithm and thus, making the performance converged to the
theoretical results.


\section{Conclusions}
In this paper, we have derived the minimum differential feedback rate for the
time-correlated Rayleigh block-fading channels considering channel
estimation error and quantization distortion. We found that the minimum
differential feedback rate is the lower bound of feedback
compression with time correlation. We also investigated the
relationship between the ergodic capacity and the feedback interval
provided the feedback-channel constraint ${C_{fb}}$ per fading
block. We found that the ergodic capacity is a monotonic convex
function on feedback intervals, and there exists an optimal feedback
interval to maximum the ergodic capacity. The simulation results of
a practical differential feedback with Lloyd's quantization
algorithm is provided to validate our theoretical results. 

\section*{Appendix A: Proof of (\ref{eq:estimation-2})}
Substituting (\ref{eq:3}) into (\ref{eq:estimation-2}), it yields
\begin{equation}\label{eq:16}
\bf{\Psi}  = \left( {1 - \frac{{\sigma _h^2}}{{\sigma _{\hat h}^2}}}
\right)\mathbf{H} - \frac{{\sigma _h^2}}{{\sigma _{\hat
h}^2}}{\mathbf{H}_e},
\end{equation}
where $\sigma _h^2$ and $\sigma _{\hat h}^2$ are the variances of
the entries of $\mathbf{H}$ and $\hat \mathbf{H}$, respectively.
Since the entries ${h_{i,j}}$ of $\mathbf{H}$, and ${h_{e,i,j}}$ of
${\mathbf{H}_e}$ are i.i.d. complex Gaussian variables, the entries
${\psi _{i,j}}$ of $\bf{\Psi}$ are also i.i.d variables. Therefore,
we only need to prove the one-dimensional model. For simplicity, the
foot labels are ignored. From (\ref {eq:16}), we can get
\begin{equation}\label{eq:17}
\psi  = \left( {1 - \frac{{\sigma _h^2}}{{\sigma _{\hat h}^2}}}
\right)h - \frac{{\sigma _h^2}}{{\sigma _{\hat h}^2}}{h_e}.
\end{equation}
From (\ref {eq:17}), as ${h_e}$ is independent on $h$ at ML cannel
estimation, the variance of $\psi$ can be calculate by
\begin{equation}\label{eq:18}
\sigma _\psi ^2 = \frac{{\sigma _h^2\left( {\sigma _{\hat h}^2 -
\sigma _h^2} \right)}}{{\sigma _{\hat h}^2}}.
\end{equation}
As a result, the distribution of $\psi$ is given by
\begin{equation}\label{eq:19}
\mathcal {C}\mathcal {N}\left( {0,\frac{{\sigma _h^2\left( {\sigma
_{\hat h}^2 - \sigma _h^2} \right)}}{{\sigma _{\hat h}^2}}} \right).
\end{equation}

In the next, we give the proof that $\psi$ is independent of $\hat
h$. As a complex Gaussian variable, $\hat h$ can be written as $\hat
h = \hat x + j \cdot \hat y$, where $\hat x$ and $\hat y$ are
$\mathcal {N}\left( {0,\frac{{\sigma _{\hat h}^2}}{2}} \right)$.
Similarly, $h$ can be written as $h = x + j \cdot y$, where $x$ and
$y$ are $\mathcal {N}\left( {0,\frac{{\sigma _h^2}}{2}} \right)$. We
then consider the conditional probability $p\left( {h|\hat h}
\right)$ when $\hat h$ is given. For the real part, the probability
can be written as
\begin{displaymath}
p\left( {x|\hat x} \right) = \frac{{p\left( {\hat x|x}
\right)p\left( x \right)}}{{p\left( {\hat x} \right)}} =
\frac{{\frac{1}{{\sqrt {\pi \left( {\sigma _{\hat h}^2 - \sigma
_h^2} \right)} }} \exp \left( { - \frac{{{{\left( {\hat x - x}
\right)}^2}}}{{\sigma _{\hat h}^2 - \sigma _h^2}}} \right)
\frac{1}{{\sqrt {\pi \sigma _h^2} }}\exp \left( { -
\frac{{{x^2}}}{{\sigma _h^2}}}
 \right)}}{{\frac{1}{{\sqrt {\pi \sigma _{\hat h}^2} }}\exp \left( { - \frac{{{x^2}}}{{\sigma _{\hat h}^2}}}
 \right)}}.
 \end{displaymath}
Therefore, we have
\begin{equation}\label{eq:20}
p\left( {x|\hat x} \right)= \frac{1}{{\sqrt {\pi \frac{{\sigma
_h^2\left( {\sigma _{\hat h}^2 - \sigma _h^2} \right)}}{{\sigma
_{\hat h}^2}}} }}\exp \left[ { - \frac{{{{\left( {x - \frac{{\sigma
_h^2}}{{\sigma _{\hat h}^2}}\hat x} \right)}^2}}}{{\frac{{\sigma
_h^2\left( {\sigma _{\hat h}^2 - \sigma _h^2} \right)}}{{\sigma
_{\hat h}^2}}}}} \right].
\end{equation}
Similarly, the imaginary part can be written as
\begin{equation}\label{eq:21}
p\left( {y|\hat y} \right) = \frac{1}{{\sqrt {\pi \frac{{\sigma
_h^2\left( {\sigma _{\hat h}^2 - \sigma _h^2} \right)}}{{\sigma
_{\hat h}^2}}} }}\exp \left[ { - \frac{{{{\left( {y - \frac{{\sigma
_h^2}}{{\sigma _{\hat h}^2}}\hat y} \right)}^2}}}{{\frac{{\sigma
_h^2\left( {\sigma _{\hat h}^2 - \sigma _h^2} \right)}}{{\sigma
_{\hat h}^2}}}}} \right].
\end{equation}

Combining (\ref {eq:20}) and (\ref {eq:21}), when $\hat h$ is given,
the conditional distribution of $h$ can be calculated as
\begin{equation}\label{eq:22}
\mathcal {C}\mathcal {N}\left( {\frac{{\sigma _h^2}}{{\sigma _{\hat
h}^2}}\hat h,\frac{{\sigma _h^2\left( {\sigma _{\hat h}^2 - \sigma
_h^2} \right)}}{{\sigma _{\hat h}^2}}} \right).
\end{equation}
Since $\psi  = h - \frac{{\sigma _h^2}}{{\sigma _{\hat h}^2}}\hat
h$, the conditional distribution of $\psi$ given $\hat h$ is given
by
\begin{equation}\label{eq:23}
\mathcal {C}\mathcal {N}\left( {0,\frac{{\sigma _h^2\left( {\sigma
_{\hat h}^2 - \sigma _h^2} \right)}}{{\sigma _{\hat h}^2}}} \right).
\end{equation}
From (\ref {eq:19}) and (\ref {eq:23}), we find that the distribution of $\psi$ is the same
regardless of whether $\hat h$ is given or not. Hence, $\psi$ is
independent of $\hat h$. Finally, the independent property between $\bf{\Psi}$ and $\hat
\mathbf{H}$ has been proved.

\section*{Appendix B: Proof of Lemma 1}

From (\ref {eq:3}), we have
\begin{equation}\label{eq:24}
{\hat h_n} = {h_n} + {h_{en}}.
\end{equation}
From (\ref {eq:2}), the one-dimensional AR(1) channel model can be
rewritten as a scalar form
\begin{equation}\label{eq:25}
{h_n} = \alpha {h_{n - 1}} + \sqrt {1 - {\alpha ^2}} {w_n}.
\end{equation}
Substituting (\ref {eq:25}) into (\ref {eq:24}) yields
\begin{equation}\label{eq:26}
\hat h_n = \left( {\alpha {h_{n - 1}} + \sqrt {1 - {\alpha ^2}}
{w_n}} \right) + {h_{en}}.
\end{equation}
From (\ref{eq:estimation-2}), we have
\begin{equation}\label{eq:27}
{h_{n - 1}} = \frac{{\sigma _h^2}}{{\sigma _{\hat h}^2}}{\hat h_{n -
1}} + {\psi _{n - 1}}.
\end{equation}
where ${\psi _n}$ is independent on ${\hat h_n}$, as proved in
Appendix A. Substituting (\ref {eq:27}) to (\ref {eq:26}) yields
\begin{equation}\label{eq:28}
\hat h_n = \alpha \left( {\frac{{\sigma _h^2}}{{\sigma _{\hat
h}^2}}{{\hat h}_{n - 1}} + {\psi _{n - 1}}} \right) + \sqrt {1 -
{\alpha ^2}} {w_n} + {h_{en}}.
\end{equation}
From (\ref{eq:CSI quantization}), we have
\begin{equation}\label{eq:29}
{\hat h_{n - 1}} = {\bar h_{n - 1}} + {e_{n - 1}}.
\end{equation}
Substituting (\ref {eq:29}) into (\ref {eq:28}), we obtain
\begin{equation}\label{eq:30}
\hat h_n = \alpha \frac{{\sigma _h^2}}{{\sigma _{\hat h}^2}}{\bar
h_{n - 1}} + \alpha \frac{{\sigma _h^2}}{{\sigma _{\hat h}^2}}{e_{n
- 1}} + \alpha {\psi _{n - 1}} + \sqrt {1 - {\alpha ^2}} {w_n} +
{h_{en}}.
\end{equation}

When ${\bar h_{n - 1}}$ is given, the conditional mutual information
can be written as
\begin{equation}\label{eq:31}
I\left( {{{\hat h}_n};{{\bar h}_n}|{{\bar h}_{n - 1}}} \right) =
h\left( {{{\hat h}_n}|{{\bar h}_{n - 1}}} \right) - h\left( {{{\hat
h}_n}|{{\bar h}_n},{{\bar h}_{n - 1}}} \right).
\end{equation}
Substituting (\ref {eq:30}) into (\ref {eq:31}), it yields
\begin{equation}\label{eq:32}
I = h\left( {\alpha \frac{{\sigma _h^2}}{{\sigma _{\hat h}^2}}{e_{n
- 1}} + \alpha {\psi _{n - 1}} + \sqrt {1 - {\alpha ^2}} {w_n} +
{h_{en}}} \right)- h\left( {{e_n}|{{\bar h}_{n - 1}}} \right).
\end{equation}
Considering the identical equation $h\left( {{e_n}|{{\bar h}_{n -
1}}} \right) \le h\left( {{e_n}} \right)$, and $h\left( {{e_n}}
\right) = h\left( {{e_{n - 1}}} \right)$, (\ref {eq:32}) can be
written as
\begin{equation}\label {eq:33}
I \ge h\left( {\alpha \frac{{\sigma _h^2}}{{\sigma _{\hat
h}^2}}{e_{n - 1}} + \alpha {\psi _{n - 1}} + \sqrt {1 - {\alpha ^2}}
{w_n} + {h_{en}}} \right)- h\left( {{e_{n - 1}}} \right).
\end{equation}
Then, (\ref {eq:33}) can be written as
\begin{equation}\label{eq:34}
I \ge h\left( {{e_{n - 1}} + \frac{{\sigma _{\hat h}^2}}{{\sigma
_h^2}}{\psi _{n - 1}} + \frac{{\sqrt {1 - {\alpha ^2}} }}{\alpha
}\frac{{\sigma _{\hat h}^2}}{{\sigma _h^2}}{w_n} + \frac{{\sigma
_{\hat h}^2}}{{\alpha \sigma _h^2}}{h_{en}}} \right) - h\left(
{{e_{n - 1}}} \right) + 2\log \left( {\alpha \frac{{\sigma
_h^2}}{{\sigma _{\hat h}^2}}} \right).
\end{equation}

As ${\bar h_{n - 1}}$, ${e_{n - 1}}$, ${\psi _{n - 1}}$, ${w_n}$ and
${h_{en}}$ are independent complex Gaussian variables, and also
mutually independent between each other, (\ref {eq:34}) can be
written as
\begin{displaymath}
I \ge I\left( {{e_{n - 1}} + \frac{{\sigma _{\hat h}^2}}{{\sigma
_h^2}}{\psi _{n - 1}} + \frac{{\sqrt {1 - {\alpha ^2}} }}{\alpha
}\frac{{\sigma _{\hat h}^2}}{{\sigma _h^2}}{w_n} + \frac{{\sigma
_{\hat h}^2}}{{\alpha \sigma _h^2}}{h_{en}};} \right. \left.
{\frac{{\sigma _{\hat h}^2}}{{\sigma _h^2}}{\psi _{n - 1}} +
\frac{{\sqrt {1 - {\alpha ^2}} }}{\alpha }\frac{{\sigma _{\hat
h}^2}}{{\sigma _h^2}}{w_n} + \frac{{\sigma _{\hat h}^2}}{{\alpha
\sigma _h^2}}{h_{en}}} \right)
\end{displaymath}
\begin{equation}\label{eq:35}
+ 2\log \left( {\alpha \frac{{\sigma _h^2}}{{\sigma _{\hat h}^2}}}
\right).
\end{equation}
According to the rate distortion theory of continuous amplitude
sources \cite{Ref:R.J-rate-distotion}, (\ref {eq:35}) achieves the
minimum value when the ${\bar h_{n - 1}}$, ${e_{n - 1}}$, ${\psi _{n
- 1}}$, ${w_n}$ and ${h_{en}}$ are independent Gaussian variables.
\begin{equation}\label{eq:36}
I \ge \log \left[ {1 + \frac{{1 - {\alpha ^2}}}{{d \cdot {\alpha
^2}}}\frac{{\sigma _{\hat h}^4}}{{\sigma _h^2}} + \frac{{\sigma
_{\hat h}^4}}{{d \cdot \sigma _h^4}}\left( {\frac{{\sigma
_h^2}}{{\sigma _{\hat h}^2}} + \frac{1}{{{\alpha ^2}}}}
\right)\left( {\sigma _{\hat h}^2 - \sigma _h^2} \right)} \right]+
2\log \left( {\alpha \frac{{\sigma _h^2}}{{\sigma _{\hat h}^2}}}
\right).
\end{equation}
From (\ref {eq:36}), we finally obtain
\begin{equation}\label{eq:37}
I \ge \log \left[ {{\alpha ^2}{{\left( {\frac{{\sigma _h^2}}{{\sigma
_{\hat h}^2}}} \right)}^2} + \frac{{\left( {1 - {\alpha ^2}}
\right)}}{d}\sigma _h^2 + \frac{{\left( {\sigma _{\hat h}^2 - \sigma
_h^2} \right)}}{d}\left( {1 + {\alpha ^2}\frac{{\sigma
_h^2}}{{\sigma _{\hat h}^2}}} \right)} \right].
\end{equation}

\section*{Appendix C: Proof of Existence of the Optimal $T$}

For simplicity, we assume $x = 2\pi {f_d}\tau $. Thus, the time
correlation can be rewritten as $\alpha  = {J_0}\left( {2\pi
{f_d}\tau } \right) = {J_0}\left( x \right)$ and the time interval
can be rewritten as $T = \tau /{t_{block}} = x/\left( {2\pi {f_d}
\cdot {t_{block}}} \right)$.
 From (\ref{eq:D-T}), the distortion $d$ can be rewritten as
 \begin{equation}\label{eq:C-d_x}
d(x) = \left( {\frac{{\sigma _h^4}}{{\sigma _{\hat h}^2}}}
\right)\left( {\frac{{\left( {1 - {2^{kx}}} \right){J_0}{{\left( x
\right)}^2}}}{{{2^{kx}} - {{\left( {\frac{{\sigma _h^2}}{{\sigma
_{\hat h}^2}}} \right)}^2}{J_0}{{\left( x \right)}^2}}}} \right) +
\sigma _{\hat h}^2.
\end{equation}
where $k = {C_{fb}}/\left( {2\pi {N_r}{N_t}{f_d} \cdot {t_{block}}}
\right)$, and $d(x)$ is a continuously differentiable function on
$x$. Then we get the first derivative $\frac{d}{{dx}}d(x)$ from
(\ref{eq:C-d_x}), we have

\begin{equation}\label{eq:D(d)}
\frac{d}{{dx}}d(x) = \frac{{{2^{kx}}\left( {\frac{{\sigma
_h^4}}{{\sigma _{\hat h}^2}}} \right)\cdot\left\{ {\left[ {2\left(
{{2^{kx}} - 1} \right)\cdot{J_1}(x) - k\ln 2\cdot\left( {{J_0}(x) -
{{\left( {\frac{{\sigma _h^2}}{{\sigma _{\hat h}^2}}}
\right)}^2}{J_0}{{(x)}^3}} \right)} \right]{J_0}(x)}
\right\}}}{{{{\left( {{2^{kx}} - {{\left( {\frac{{\sigma
_h^2}}{{\sigma _{\hat h}^2}}} \right)}^2}{J_0}{{\left( x
\right)}^2}} \right)}^2}}}.
\end{equation}
where ${J_1}\left( x \right) =  - \frac{d}{{dx}}{J_0}\left( x
\right)$ in [17], where ${J_n}\left( x \right)$ is a first kind
$n$-order Bessel function.

When $x \to 0$, there are ${J_0}(x) \to 1$ and ${J_1}(x) \to 0$.
Thus, the first derivative of $d(x)$ is
\begin{equation}\label{eq:less than o}
\frac{d}{{dx}}d(x){|_{x \to 0}} =  - \left\{ {\frac{{\frac{{\sigma
_h^4}}{{\sigma _{\hat h}^2}}\cdot\left[ {k\ln 2\cdot\left( {1 -
{{\left( {\frac{{\sigma _h^2}}{{\sigma _{\hat h}^2}}} \right)}^2}}
\right)} \right]}}{{{{\left( {1 - {{\left( {\frac{{\sigma
_h^2}}{{\sigma _{\hat h}^2}}} \right)}^2}} \right)}^2}}}} \right\} <
0.
\end{equation}

However, when $x = \frac{3}{2}$, since ${J_1}\left( {\frac{3}{2}}
\right)
> {J_0}\left( {\frac{3}{2}} \right)$, we have
\begin{displaymath}
\frac{d}{{dx}}d(x){|_{x = \frac{3}{2}}} >
\frac{{{2^{\frac{3}{2}k}}\left( {\frac{{\sigma _h^4}}{{\sigma _{\hat
h}^2}}} \right)\cdot\left\{ {\left[ {2\left( {{2^{\frac{3}{2}k}} -
1} \right) - k\ln 2\cdot\left( {1 - {{\left( {\frac{{\sigma
_h^2}}{{\sigma _{\hat h}^2}}} \right)}^2}{J_0}{{\left( {\frac{3}{2}}
\right)}^2}} \right)} \right]{J_0}{{\left( {\frac{3}{2}}
\right)}^2}} \right\}}}{{{{\left( {{2^{\frac{3}{2}k}} - {{\left(
{\frac{{\sigma _h^2}}{{\sigma _{\hat h}^2}}}
\right)}^2}{J_0}{{\left( {\frac{3}{2}} \right)}^2}} \right)}^2}}}
\end{displaymath}
\begin{equation}\label{eq:inequlity-1}
> \frac{{{2^{\frac{3}{2}k}}\left( {\frac{{\sigma _h^4}}{{\sigma
_{\hat h}^2}}} \right)\cdot\left\{ {\left[ {2\left(
{{2^{\frac{3}{2}k}} - 1} \right) - k\ln 2} \right]{J_0}{{\left(
{\frac{3}{2}} \right)}^2}} \right\}}}{{{{\left( {{2^{\frac{3}{2}k}}
- {{\left( {\frac{{\sigma _h^2}}{{\sigma _{\hat h}^2}}}
\right)}^2}{J_0}{{\left( {\frac{3}{2}} \right)}^2}} \right)}^2}}}
\end{equation}
Considering the inequality ${2^{\frac{3}{2}k}} - 1 > \frac{3}{2}\ln
2 \cdot k{2^{\frac{3}{2}k}}$, we have
\begin{equation}\label{eq:inequality}
2\left( {{2^{\frac{3}{2}k}} - 1} \right) - k\ln 2 > k\ln 2\left( {3
\cdot {2^{\frac{3}{2}k}} - 1} \right) > 0.
\end{equation}
Substituting (\ref{eq:inequality}) to (\ref{eq:inequlity-1}), we
have
\begin{equation}\label{eq:more than o}
\frac{d}{{dx}}d(x){|_{x = \frac{3}{2}}} > 0.
\end{equation}

As $\frac{d}{{dx}}d(x)$ is a continuous function on $x$, combining
(\ref{eq:less than o}) and (\ref{eq:more than o}), we can easily
obtain there exists a $x$ to make $\frac{d}{{dx}}d(x) = 0$ when $0 <
{x_{opt}} < \frac{3}{2}$. Thus, the existence of the optimal
${T_{opt}} = {x_{opt}}/\left( {2\pi {f_d}\cdot{t_{block}}} \right)$
is proved.


\pagebreak

\clearpage

\clearpage

\clearpage

\clearpage


\end{document}